\documentclass[twocolumn,superscriptaddress,aps]{revtex4}
\usepackage{epsfig}
\usepackage{graphicx}
\usepackage{bm}
\begin{document}
\title{Cyclic permutation-time symmetric structure with coupled gain-loss microcavities}
\author{Bing He}
\affiliation{Department of Physics, University of Arkansas, Fayetteville, Arkansas 72701, USA}
\author{Liu Yang }
\affiliation{Department of Physics, University of Arkansas, Fayetteville, Arkansas 72701, USA}
\affiliation{ College of Physics, Jilin University, Changchun 130012, China   }
\author{Zhaoyang Zhang }
\affiliation{Department of Physics, University of Arkansas, Fayetteville, Arkansas 72701, USA}
\affiliation{ Key Laboratory for Physical Electronics and Devices of the Ministry of Education, Xi'an Jiaotong University, Xi'an 
710049, China}
\author{Min Xiao}
\affiliation{Department of Physics, University of Arkansas, Fayetteville, Arkansas 72701, USA}
\affiliation{National Laboratory of Solid State Microstructures and School of Physics, Nanjing University, Nanjing 210093, China}

\begin{abstract}
We study the coupled even number of microcavities with the balanced gain and loss between any pair of their neighboring components. The effective non-Hermitian Hamiltonian for such structure has the cyclic permutation-time symmetry with respect to the cavity modes, and this symmetry determines the patterns of the dynamical evolutions of the cavity modes. The systems also have multiple exceptional points for the degeneracy of the existing supermodes, exhibiting the ``phase transition'' of system dynamics across these exceptional points. 
We illustrate the quantum dynamical properties of the systems with the evolutions of cavity photon numbers and correlation functions. Moreover, we demonstrate the effects of the quantum noises accompanying the amplification and dissipation of the cavity modes. The reciprocal light transportation predicted with the effective non-Hermitian models for the similar couplers is violated by the unavoidable quantum noises.
\end{abstract}
\maketitle

\section{introduction}
Coupled photonic structures exhibit rich physics and have wide applications. One category that has attracted wide research interest 
in the past decade is with properly balanced gain-loss profiles for the coupling optical modes, as it is regarded as a realization of optical analogue for parity-time (PT)-symmetric quantum mechanics \cite{bender, bender2}. These proposed systems include optical waveguides \cite{w1, o1, o2}, optical lattices \cite{o1,o2,o3}, microcavities \cite{m1,m2}, and others \cite{oth, oth2}.
So far most of the researches on these systems are concerned with the properties such as light transportation, 
and some of the interesting features of the systems have been demonstrated by the experiments with different setups \cite{ex1, ex2, ex3, ex4, ex5}. An important character of a PT-symmetric system is the existence of the threshold known as exceptional point \cite{e1,e2}, across which the optical modes undergo a ``phase transition'' between periodic oscillation and exponential growth/decay. This feature can be applied to control the transmitted light intensity and enhance the added nonlinearity for generating nonclassical light field states with PT systems \cite{bhe}.

In addition to the above mentioned properties, the quantum features of linear PT systems have received attention recently \cite{m2, oth2, arga}. The quantum dynamics of these systems should be well understood in view of the potential applications. On the one hand, the dynamics of these systems has been studied with PT-symmetric non-Hermitian Hamiltonians, which have real and complex eigenvalues, respectively, across the exceptional point where there is the unit ratio of amplification (dissipation) rate over coupling strength. 
The non-Hermitian system Hamiltonians lead to unique properties of quantum state transfer in coupled PT-symmetric arrays \cite{m2, oth2}. 
On the other hand, as in any open quantum system, quantum noises coexisting with the gain and loss of the optical modes will inevitably affect the dynamics of the systems. The associated noises destroy the nonclassicality of quantum light sent into a linear PT-symmetric coupler \cite{arga}. The external quantum noise driving fields, as well as the amplification and dissipation of light fields, can be described in terms of a stochastic Hermitian Hamiltonian \cite{book}. A clearer understanding of the possible differences between the two pictures is meaningful to further researches on the quantum aspect of similar coupled gain-loss systems. 

In this paper we propose a generalized structure from PT symmetry. The concerned linear quantum systems have alternately distributed equal gain and loss for their neighboring microcavity modes, so that their dynamical equations without external driving terms are invariant under cyclic permutation operations plus time reversal operation for the coupling modes. One consequence of this cyclic permutation-time (CPT) symmetry is the existence of multiple exceptional points.
We will illustrate the dynamical behaviors of the systems across these exceptional points. The other purpose of the current work is 
to understand the effects of quantum noises on this structure. The light transportation through a CPT system is compared for both non-Hermitian model and stochastic Hermitian Hamiltonian including quantum noises. We show that the corrections from the noises are considerable to such linear dynamical systems if the transmitted light is not so strong. 

The rest of the paper is organized as follows. In Sec. II we provide the description of the CPT systems with the non-Hermitian effective Hamiltonian, as well as with the stochastic Hermitian Hamiltonian including the quantum noises. The quantum dynamics of the systems is discussed with an example in Sec. III; here we illustrate the evolution of cavity photon numbers and a kind of quantum correlation functions in the different regimes across the exceptional points. In Sec. IV we study the quantum noise effects to show their influence on the light transportation through a CPT system. The difference between the non-Hermitian Hamiltonian approach and the stochastic Hermitian Hamiltonian approach will be discussed in this section. Finally we conclude the current study in Sec. V.   

\begin{figure}[t!]
\vspace{0cm}
\centering
\epsfig{file=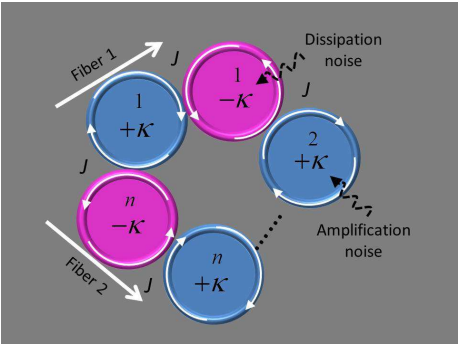,width=0.7\linewidth,clip=} 
{\vspace{-0.2cm}\caption{(color online) A schematic of CPT systems with even number of microcavities, which carry the balanced gain and loss (at the rate $\kappa$) for each pair of the neighboring components. These cavities are coupled with each other by an adjustable strength $J$. In the non-Hermitian model for the system, the light transported from Fiber 1 to Fiber 2 has the same intensity 
as in the reverse process. Here we also consider the quantum noises as the random driving fields into the cavities.  \label{fig: 1}  }}
\vspace{0cm}
\end{figure}

\section{CPT systems as non-Hermtian model and under quantum noises  }

A generalized CPT system from the PT-symmetric ones is illustrated in Fig. \ref{fig: 1}. Here the microcavities carry the balanced gain and loss at the rate $\kappa$ for each pair of neighboring components, which couple with the strength $J$. Without loss of generality 
we let the external coherent driving field act on one of the gain cavities. This configuration of even numbers of microcavities with the balanced gain and loss possesses interesting patterns for the involved dynamical supermodes illustrated below, which do not exist in coupled systems of 
odd numbers of cavities. We use the mode operator $\hat{a}_i$ ($i=1,2,\cdots, n$) to denote those of the amplification cavities, while those of the dissipation ones are denoted as $\hat{b}_i$. 
They satisfy the commutation relation $[\hat{c}_i,\hat{c}_j^\dagger]=\delta_{ij}$ for $\hat{c}_i=\hat{a}_i,\hat{b}_i$. 
The dynamical equation for the system reads ($\hbar\equiv 1$)
\begin{eqnarray}
i\frac{d}{dt}\vec{c}(t)&=&[M]\vec{c}(t)+\vec{d}(t),\nonumber\\
 ~[M]&=& \left(
\begin{array}
[c]{cccccc}
i\kappa & J &  &  & & J \\
J & -i\kappa & &  &\\
\cdots & \cdots &\cdots &\cdots &\cdots & \cdots\\
    & & J & i\kappa & J & \\
\cdots & \cdots &\cdots &\cdots &\cdots &\cdots \\
J &   & &  & J & -i\kappa    \\
\end{array}
\right)
\label{dynamics}
\end{eqnarray}
in the interaction picture with respect to the self oscillation Hamiltonian of the microcavities,
where $\vec{c}(t)=(\hat{a}_1(t),\hat{b}_1(t),\cdots, \hat{a}_n(t),\hat{b}_n(t))^T$, and $\vec{d}(t)$ denotes the external drives.
The external coherent driving filed is into the first gain cavity of such linear systems, i.e. $d_1=iE e^{i\Delta t}$, without loss of generality, where $\Delta$ is the detuning of the driving laser frequency from the intrinsic cavity frequency and $E$ the amplitude. 
Similar to the non-Hermitian Hamiltonians in \cite{m2, oth2}, the dynamical process in Eq. (\ref{dynamics}) corresponds to the effective Hamiltonian
\begin{eqnarray}
H_{eff}&=&i\kappa \hat{a}_1^\dagger\hat{a}_1+\cdots+i\kappa \hat{a}_n^\dagger\hat{a}_n\nonumber\\
&-& i\kappa \hat{b}_1^\dagger\hat{b}_1-\cdots-i\kappa \hat{b}_n^\dagger\hat{b}_n\nonumber\\
&+& J(\hat{a}_1\hat{b}_1^\dagger+\hat{a}_1^\dagger\hat{b}_1)+\cdots+ J(\hat{a}_1\hat{b}_n^\dagger+\hat{a}_1^\dagger\hat{b}_n)\nonumber\\
&+&i E(\hat{a}^\dagger_1 e^{i\Delta t}-\hat{a}_1 e^{-i\Delta t}).
\label{nh}
\end{eqnarray}
Without the external driving field, this Hamiltonian is invariant under the cyclic permutation operation 
${\cal CP}(\hat{a}_1,\hat{b}_1,\cdots,\hat{a}_n,\hat{b}_n)=(\hat{b}_1, \hat{a}_2,\cdots, \hat{b}_n, \hat{a}_1)$ plus the time reversal operation ${\cal T}i=-i$, hence the symmetry of the dynamical equation Eq. (\ref{dynamics}). One consequence of such symmetry is the reciprocal light transportation through the systems, if added with an external driving field. This inherits the same property of a linear PT-symmetric coupler, i.e. the light driven into the active (gain) cavity and going out of the passive (loss) one has the same intensity as in the reverse process (into the passive one and out of the active one). 

In reality the amplification and dissipation of the cavity modes are also accompanied by the quantum noises. These noises act as the additional random driving fields to the coherent one in Eq. (\ref{dynamics}), and then the total drive vector in the equation takes the 
form $\vec{d}(t)=(iEe^{i\Delta t},0,\cdots,0)^T+i\sqrt{2\kappa}(\hat{\xi}_1^\dagger(t),\hat{\eta}_1(t),\cdots,\hat{\xi}_n^\dagger(t),\hat{\eta}_n(t))^T$.
The amplification noise operator $\hat{\xi}_i(t)$ and the dissipation noise operator $\hat{\eta}_i(t)$ satisfy the correlation relation $\langle \hat{\xi}_i(t)\hat{\xi}_i^\dagger(t')\rangle=\langle \hat{\eta}_i(t)\hat{\eta}_i^\dagger(t')\rangle=\delta(t-t')$. The noises together with the cavity gains will make contribution to the output photon numbers, known as spontaneous photon generation \cite{arga}.
The gain and loss of the cavity modes in Eq. (\ref{dynamics}) can be described by their coupling with the associated reservoirs, so the process is due to a stochastic Hamiltonian \cite{book} (the notation for the amplification noise operator follows that in \cite{bhe})
\begin{eqnarray}
H&=& J(\hat{a}_1\hat{b}_1^\dagger+\hat{a}_1^\dagger\hat{b}_1)+\cdots+ J(\hat{a}_1\hat{b}_n^\dagger+\hat{a}_1^\dagger\hat{b}_n)\nonumber\\
&+& i\sqrt{2\kappa}\{(\hat{a}_1^\dagger \hat{\xi}_1^\dagger(t)-\hat{a}_1 \hat{\xi}_1(t))+\cdots+(\hat{a}_n^\dagger \hat{\xi}_n^\dagger(t)-\hat{a}_n \hat{\xi}_n(t))\}\nonumber\\
&+&i E(\hat{a}^\dagger_1 e^{i\Delta t}-\hat{a}_1 e^{-i\Delta t})\nonumber\\
&+& i\sqrt{2\kappa}\{(\hat{b}^\dagger_1 \hat{\eta}_1(t)-\hat{b}_1 \hat{\eta}^\dagger_1(t))+\cdots+(\hat{b}_n^\dagger \hat{\eta}_n(t)-\hat{b}_n \hat{\eta}^\dagger_n(t))\}.\nonumber\\
\label{h}
\end{eqnarray}
Obviously the CPT symmetry will be broken for this stochastic Hamiltonian (excluding the coherent drive term). The consequence of the symmetry breaking for the linear couplers in Fig. 1 is one of the main points we will discuss below. 

\section{dynamics across multiple exceptional points}

\subsection{existence of multiple exceptional points}
The dynamics of the CPT systems is completely determined by the matrix $[M]$ in Eq. (\ref{dynamics}). Unlike a Hermitian matrix this CPT-symmetric matrix does not have a real spectrum decomposition in terms of the orthonormal eigenvectors as the supermodes. However, various interesting properties arise from its non-Hermitianity. 

\begin{table}[t!]%
\begin{tabular}
[c]{|c|c|}\hline
       & ($\lambda, m$)  \\\hline 
$N=2$ & $(-\sqrt{\kappa^2-J^2},1)$, $(\sqrt{\kappa^2-J^2},1)$ \\\hline
$N=4$ & $(-\kappa,1)$,$(\kappa,1)$, $(-\sqrt{\kappa^2-4J^2},1)$, $(\sqrt{\kappa^2-4J^2},1)$ \\\hline
$N=6$ & $(-\sqrt{\kappa^2-J^2},2)$, $(\sqrt{\kappa^2-J^2},2)$, $(-\sqrt{\kappa^2-4J^2},1)$,  \\\hline
& $(\sqrt{\kappa^2-4J^2},1)$ \\\hline
$N=8$ &    $(-\kappa,1)$,$(\kappa,1)$,  $(-\sqrt{\kappa^2-2J^2},2)$, $(\sqrt{\kappa^2-2J^2},2)$,  \\\hline
 & $(-\sqrt{\kappa^2-4J^2},1)$, $(\sqrt{\kappa^2-4J^2},1)$ \\\hline
 $N=10$ & $(-\sqrt{\kappa^2-(\frac{3-\sqrt{5}}{2})J^2},2)$, $(\sqrt{\kappa^2-(\frac{3-\sqrt{5}}{2})J^2},2)$, \\\hline
  & $(-\sqrt{\kappa^2-(\frac{3+\sqrt{5}}{2})J^2},2)$, $(\sqrt{\kappa^2-(\frac{3+\sqrt{5}}{2})J^2},2)$, \\\hline
  &  $(-\sqrt{\kappa^2-4J^2},1)$, $(\sqrt{\kappa^2-4J^2},1)$ \\\hline
  $N=12$ & $(-\kappa,1)$,$(\kappa,1)$, $(-\sqrt{\kappa^2-J^2},2)$, $(\sqrt{\kappa^2-J^2},2)$,\\\hline
  & $(-\sqrt{\kappa^2-3J^2},2)$, $(\sqrt{\kappa^2-3J^2},2)$,\\\hline
  &  $(-\sqrt{\kappa^2-4J^2},1)$, $(\sqrt{\kappa^2-4J^2},1)$ \\\hline
\end{tabular}
\caption{Supermode spectra for CPT systems. Here we list the eigenvalues and their degeneracy of the matrices $-i[M]$ for those with small cavity number $N$. $\lambda$ denotes the eigenvalues, and $m$ is the degeneracy of the corresponding eigenvalue.}
\vspace{-0.5cm}
\end{table}

The spectrum and supermodes of the matrix $[M]$ have regular patterns with the growth of the cavity number $N$, though they do not have the exact formulas. We list their proper forms in Tab. I for some CPT systems with small cavity number $N$. The eigenvalues of $-i[M]$ appear in pairs of opposite signs. Their distribution pattern is perceivable; the single eigenvalue pair with the values $\pm\kappa$ appears periodically with the number $N$, and all other ones except for this pair and another one have the double degeneracy which affects the forms of the corresponding supermodes (eigenvectors). The single eigenvalue pair $\pm\sqrt{\kappa^2-J^2}$ for $N=2$ is the one that exists in all linear PT-symmetric systems, and this pair of eigenvalues degenerate at $\kappa=J$ as the exceptional point. For the higher numbers $N$, there are more such points at which the pairs of eigenvalues will degenerate further into the single ones. We call them the exceptional points for the CPT systems. For instance, two exceptional points, $\kappa=J$ and $\kappa=2J$, exist for the CPT system of $N=6$. Across an exceptional point by adjusting the ratio $J/\kappa$, one pair of the eigenvalues takes the transition between the real and imaginary values, thus giving rise to different dynamical behaviors of the system. The dynamical evolution of the CPT systems can be therefore very different from the unitary evolution of a closed quantum system.

\subsection{evolution of observables}
The cavity modes evolve according to Eq. (\ref{dynamics}) to the following:
\begin{eqnarray}
c_i(t)&=&\sum_j[e^{-i[M]t}]_{ij}c_j(0)\nonumber\\
& -&i\int_0^t d\tau\sum_j [e^{-i[M](t-\tau)}]_{ij}d_j(\tau),
\label{linear}
\end{eqnarray}
where $[\cdots]_{ij}$ denote the matrix elements, and $\vec{c}(0)=(\hat{a}_1,\hat{b}_1,\cdots, \hat{a}_n,\hat{b}_n)^T$.
The interesting features of this linear dynamical evolution come from the CPT symmetry of the dynamical matrix $[M]$.
The CPT-symmetric matrix can be diagonalized as $-i[M]= [K][\Lambda][K]^{-1}$ 
(the diagonal elements of $[\Lambda]$ are the eigenvalues in Tab. I), implying $e^{-i[M]t}=[K]e^{[\Lambda]t}[K]^{-1}$. Due to the double degeneracy of the eigenvalues shown in Tab. 1, the expectation values of at least two amplification (dissipation) cavity modes will evolve in the same way for any ratio $J/\kappa$, given an external driving field into any of the coupled cavities. Take the system of $N=6$ for example, the averages $\langle \hat{a}_2(t)\rangle$ and $\langle\hat{a}_3(t)\rangle$, as well as $\langle \hat{b}_1(t)\rangle$ and $\langle\hat{b}_3(t)\rangle$, are the same under a coherent driving field into the first amplification cavity.

The dynamics across the different exceptional points should be understood for the CPT systems. 
Two types of information, the control on the transmitted light intensity with the system parameters and the possible correlations between the output modes from the individual microcavities, are useful to their potential applications.  
The former is straightforwardly quantified by the cavity photon evolving inside the microcavities and, 
for the latter, we are concerned with the phase coherence between the cavity modes $\hat{c}_i$ here. The phase coherence, as an important quantum correlation, can be measured by the average contrast \cite{c1,c2,c3}  
\begin{eqnarray}
{\cal C}_{c_ic_j}(t)=\frac{2|\langle \hat{c}^\dagger_i\hat{c}_j\rangle|}{\langle \hat{c}_i^\dagger\hat{c}_i\rangle+\langle \hat{c}_j^\dagger\hat{c}_j\rangle}
\end{eqnarray}
in interference experiments, where $\hat{c}_i=\hat{a}_i$ or $\hat{b}_i$. From the input-output relation, the measured output modes $\hat{d}_i$ in experiment are related to the cavity modes as $\hat{d}_i=\sqrt{2\gamma}\hat{c}_i$, where $\gamma$ is the coupling rate between the cavities and the optical fibers for outputting the light. There is an extra term for the cavity into which the external driving field is added. 

\begin{figure}[t!]
\vspace{0cm}
\centering
\epsfig{file=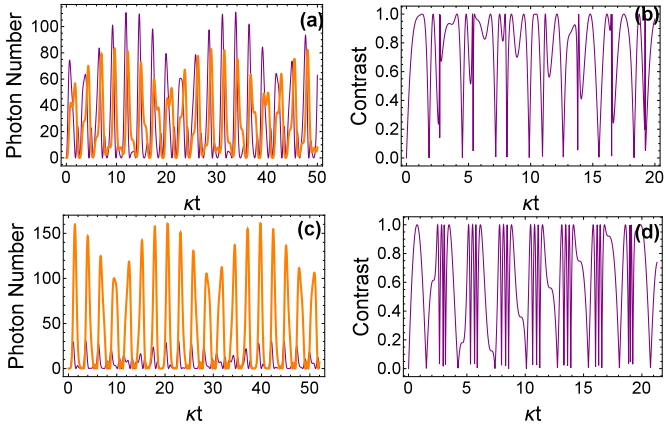,width=1.05\linewidth,clip=} 
{\vspace{-0.2cm}\caption{(color online) Evolution of photon numbers and average contrast in the regime $\kappa < J$. Here the plots are obtained with $J/\kappa=2.5$, $E/\kappa=20$ and $\Delta=0$. The initial quantum state of the system is assumed to be a vacuum state. 
(a) illustrates the evolution of $\langle \hat{a}_1^\dagger\hat{a}_1\rangle$ (the thinner purple curve) and $\langle \hat{b}_1^\dagger\hat{b}_1\rangle$ (the thicker orange curve); (b) is about the evolution of the average contrast ${\cal C}_{a_1 b_1}$ between the two modes. (c) and (d) illustrate the corresponding quantities for the modes $\hat{a}_2$ and $\hat{b}_2$. \label{fig: 2}  }}
\vspace{0cm}
\end{figure}

Some of the CPT systems with one pair of supermodes of the eigenvalue $\lambda=\pm\kappa$ (see Tab. I) are not well tunable, because the exponentially growing one of them (independent of the system parameter $J$) will quickly dominate the dynamical evolutions of the observables. Here we demonstrate the evolution patterns of the cavity photon numbers and average contrasts between the cavity modes with the example of $N=6$, which has two exceptional points at $\kappa=J$ and $\kappa=2J$, respectively. We assume that the amplification/dissipation rate $\kappa$ in the system is fixed, and the coupling strength $J$ between the cavities can be adjusted. 
We start from the strong coupling regime $J>\kappa$. Then all eigenvalues for the matrix $-i[M]$ are imaginary. In this regime the eigenvalues of the effective non-Hermitian Hamiltonian in Eq. (\ref{nh}) are all real, corresponding to the commonly known PT-symmetric regime for $N=2$. The cavity modes are simply oscillating there. Consequently the two quantities mentioned above evolve as the quasi periodic functions shown in Fig. \ref{fig: 2}. Compared with the PT-symmetric systems (the special case of $N=2$ in our generalization), more harmonic modes contribute to these quasi periodic evolutions for a high cavity number $N$. For brevity we only pick out two pairs of cavity modes to illustrate the evolution of the concerned quantities with time, and those for the other modes are similar. The extreme situation in this regime is the strong coupling limit $J\gg \kappa$, where the system behaves like a beam-splitter array used for photonic quantum information processing (see, e.g. \cite{b1,b2,b3}). In this regime with $J>\kappa$, the pure transportation of light from the externally driven cavity can dominantly determine the photon numbers in other cavities, and the field intensities in them are the results of the effective interferences from two identical ``sources" (in the driven cavity) respectively transmitting light along the clockwise and counterclockwise path in Fig. 1, which are formed by a beam-splitter type coupling between the cavities. The diagonal cavity to the driven one is under a perfect constructive interference because the photons are transported along two symmetric paths to it. In our example of driving the first amplification cavity, the second dissipation cavity as its diagonal unit therefore has a higher intensity than its neighboring amplification ones [see Fig. \ref{fig: 2}(c)], as the photons are transported along two paths on which there is a symmetry with $\langle \hat{b}^{\dagger}_1\hat{b}_1(t)\rangle=\langle \hat{b}^{\dagger}_3\hat{b}_3(t)\rangle$ and $\langle \hat{a}^{\dagger}_2\hat{a}_2(t)\rangle=\langle \hat{a}^{\dagger}_3\hat{a}_3(t)\rangle$. This phenomenon is actually a consequence of the CPT symmetry.

\begin{figure}[t!]
\vspace{0cm}
\centering
\epsfig{file=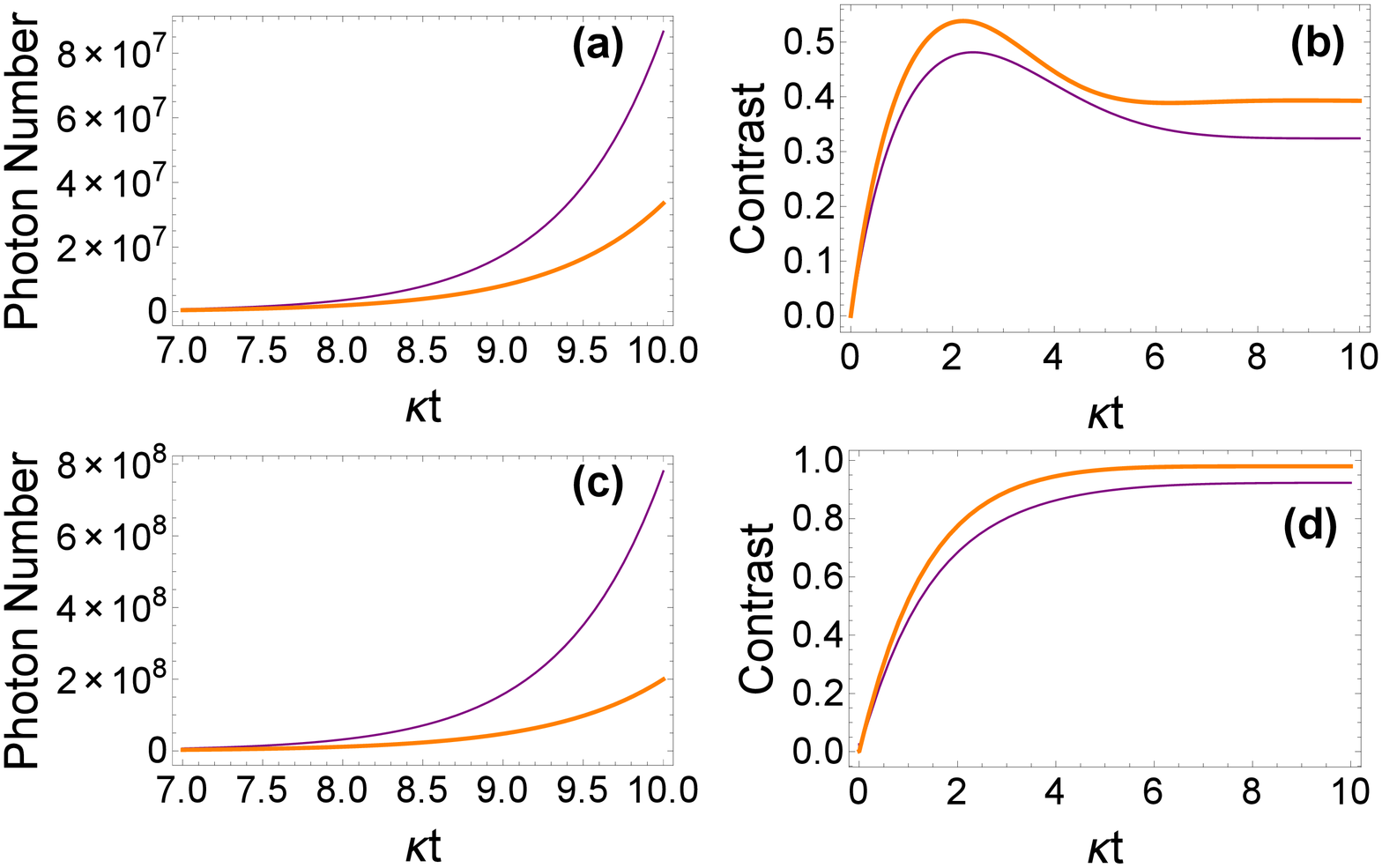,width=1.05\linewidth,clip=} 
{\vspace{-0.2cm}\caption{ (color online) Evolution of photon numbers and average contrast in the regime $J<\kappa < 2J$. In (a) the thinner (purple) solid curve stands for the evolution of $\langle \hat{b}_1^\dagger\hat{b}_1\rangle$ with $J/\kappa=0.6$ and the thicker (orange) one stands for that of $\langle \hat{b}_1^\dagger\hat{b}_1\rangle$ with $J/\kappa=0.7$. In (b) the curves stand for the evolution of the average contrast between the modes $\hat{a}_1$ and $\hat{b}_1$ for the ratios in (a), respectively. Correspondingly the plots in (c) and (d) are about the evolution of $\langle \hat{a}_2^\dagger\hat{a}_2\rangle$ [with the same $J/\kappa$ ratios as in (a)] and the average contrast between the modes $\hat{a}_2$ and $\hat{b}_2$, respectively. All other parameters are the same as in 
Fig. \ref{fig: 2}. \label{fig: 3}}}
\vspace{-0.5cm}
\end{figure}

\begin{figure}[b!]
\vspace{0cm}
\centering
\epsfig{file=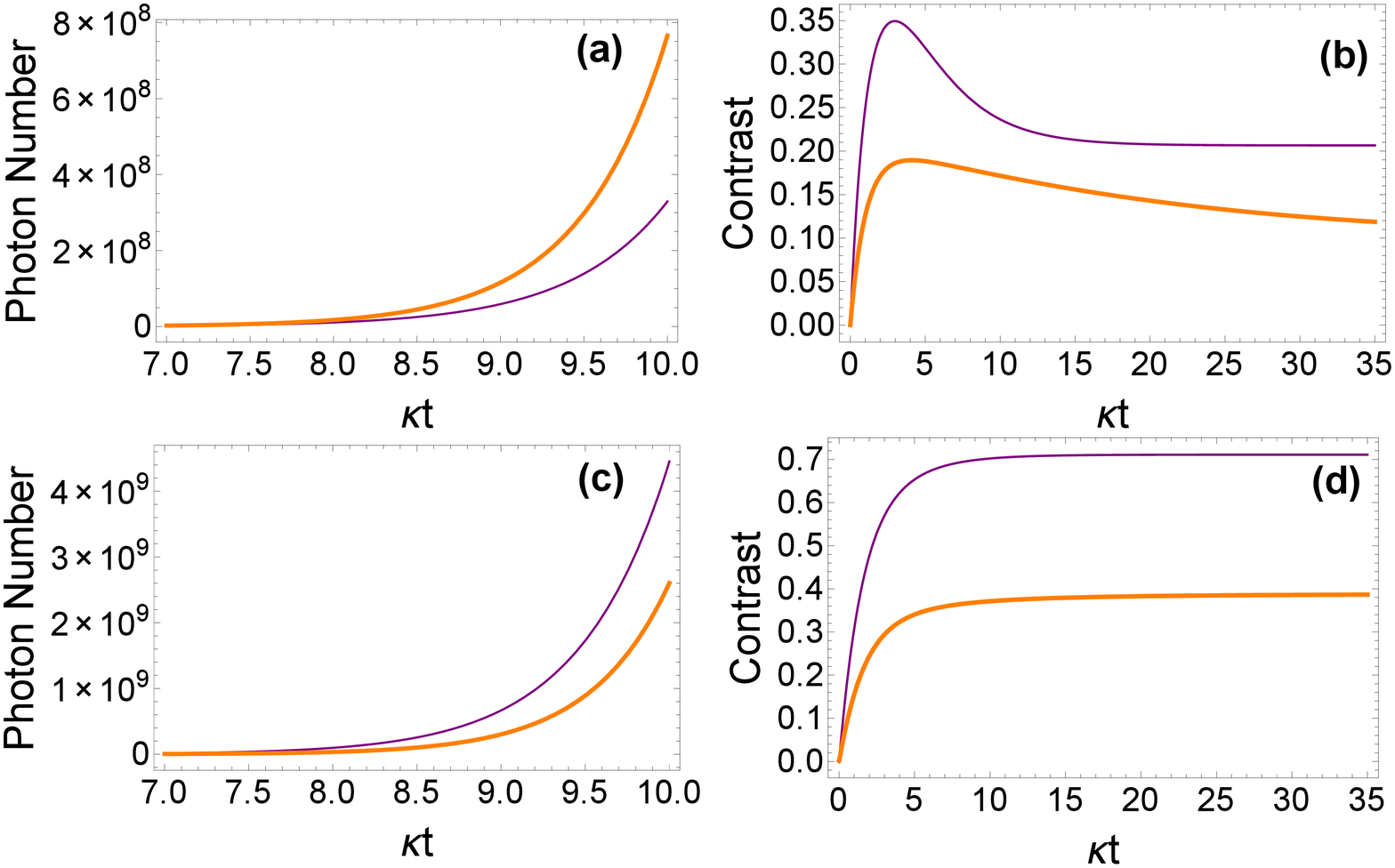,width=1.05\linewidth,clip=} 
{\vspace{-0.2cm}\caption{ (color online) Evolution of photon numbers and average contrast in the regime $\kappa > 2J$. All plots here describe the same quantities as in Fig. \ref {fig: 3}, except that the thinner (purple) curves now stands for the situation with $J/\kappa=0.4$ and the thicker (orange) ones for $J/\kappa=0.2$. \label{fig: 4} }}
\vspace{-0cm}
\end{figure}

Next we lower the coupling strength such that the system enters the regime $J<\kappa < 2J$ between the two exceptional points. 
The evolution of the cavity photon numbers and the average contrast between the cavity modes are shown in Fig. \ref{fig: 3}. In contrast to the situations in Fig. \ref{fig: 2}, the photon numbers significantly grow up with time, given the same external driving field as in 
Fig. \ref{fig: 2}. It is due to the transition of two oscillating supermodes (of the eigenvalue value $\lambda=\sqrt{\kappa^2-J^2}$ with the double degeneracy) to the exponentially increasing ones. These two supermodes will dominate in the end. As the consequence, the average contrasts in Fig. \ref{fig: 3} will become stable with time. Comparing the results in Fig. \ref{fig: 3}(a) and \ref{fig: 3} (c), one sees that the photon numbers in the amplification cavities begin to surpass those in the dissipation ones. 
As the coupling strength $J$ is reduced, the photon numbers in all cavities will steadily increase instead. The change of the average contrast with the coupling $J$ shows the opposite tendency, due to the different growth paces for the correlation $|\langle \hat{c}^\dagger_i\hat{c}_j\rangle|$ 
($i\neq j$) and the photon numbers.

\begin{figure}[b!]
\vspace{0cm}
\centering
\epsfig{file=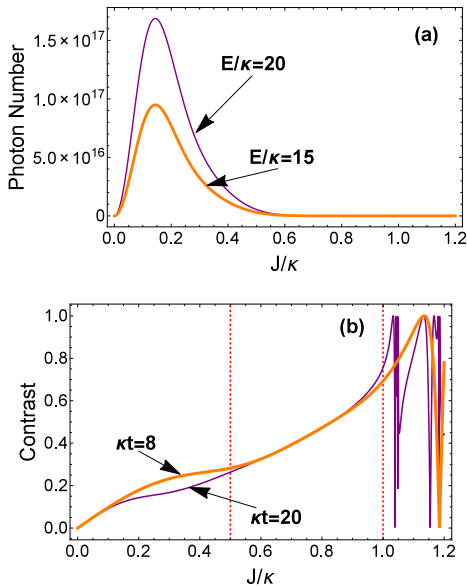,width=0.77\linewidth,clip=} 
{\vspace{-0.2cm}\caption{(color online) Comparison between the different dynamical regimes for a CPT system of $N=6$. (a) shows the 
photon number $\langle\hat{b}^\dagger_1\hat{b}_1\rangle$ distributing with the parameter $J/\kappa$ at the time $\kappa t=20$, when the supermodes of the largest eigenvalue have dominated the system evolution. (b) is the corresponding distributions of 
the average contrast ${\cal C}_{a_1b_1}$ at two different moments, under a driving field of $E/\kappa=20$. Here we mark the two boundaries of the three regimes with the dashed vertical lines. The photon numbers in other cavities and the average contrasts between other modes have the similar distributions. \label{fig: 5} }}
\vspace{0cm}
\end{figure}

\begin{figure*}[t!]
\includegraphics[width=0.9\textwidth, clip]{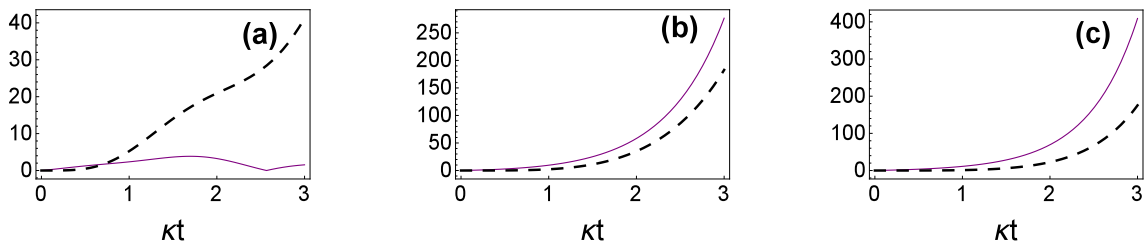}
\caption{ (color online) Difference between transported photon numbers along the mutually reverse paths. The solid curves stand for the transported photon number differences, which are proportional to the difference in the measured light powers. Here we illustrate the quantity 
$|\langle\hat{b}_3^\dagger\hat{b}_3\rangle_F -\langle \hat{a}_1^\dagger\hat{a}_1\rangle_B|$, where $\langle\hat{b}_3^\dagger\hat{b}_3\rangle_F$ is the photon number of the third dissipation cavity, which is found in the forward transmission pattern with the coherent drive on the first amplification cavity, and $\langle \hat{a}_1^\dagger\hat{a}_1\rangle_B$ is the photon number in the first amplification cavity due to the backward transmission pattern of driving the third dissipation cavity. Such difference comes from the amplification noises. The dashed curves represent the constant forwardly and backwardly transported photon numbers purely due to a coherent driving field with the intensity $E/\kappa=5$, which are found with the non-Hermitian model. (a), (b) and (c) 
are about the ratio $J/\kappa=1.2$, $0.6$ and $0.4$, respectively. \label{fig: 6}}  
\vspace{0cm}
\end{figure*}

After the coupling strength $J$ is lowered further to tune the system into the regime of $\kappa>2J$, the evolutions of the concerned quantities illustrated in Fig. \ref{fig: 4} still behave similarly to those in Fig. \ref{fig: 3}, but show the significant difference in their quantitative values (in Fig. \ref{fig: 4} we only show the beginning stage for the photon numbers to significantly deviate from those in Fig. \ref{fig: 3}). A main character in this regime is the saturation of photon amplification with the decrease of the coupling intensity $J$; see Fig. \ref{fig: 5}(a). The photon number growth rate determined by the factors $e^{\sqrt{\kappa^2-J^2}(t-\tau)}$ and $e^{\sqrt{\kappa^2-4J^2}(t-\tau)}$ from the exponentially increasing modes gets larger for a smaller $J$. However, the smaller $J$ will reduce the amount of photons transported from the driven cavity to the rest ones. These two competing tendencies result in the 
maximum cavity photon numbers at a coupling strength $J$ inside the regime. In dynamics this regime differs from the other one illustrated in Fig. \ref{fig: 3} by having one more supermode 
($\lambda=\sqrt{\kappa^2-4J^2}$) to undergo a transition from periodically oscillating to exponentially growing. 
The distinction of the two regimes possessing growing supermodes can be better seen in 
Fig. \ref{fig: 5}(b), which shows an average contrast distribution after a short and longer period of time. 
Though the supermodes with the largest eigenvalue $\lambda=\sqrt{\kappa^2-J^2}$ will dominate nonetheless, the extra growing supermode in the regime $\kappa>2J$ considerably affects the transient behavior of the correlation function, making it take longer time to become stable. In contrast, the correlation function as the average contrast quickly stabilizes in the regime $J<\kappa<2J$, causing the two distribution curves in Fig. \ref{fig: 5}(b) mostly overlapped in this middle regime. It should be noted that, at any moment, all observables distribute continuously across the exceptional points as seen from Fig. \ref{fig: 5}. This is totally different from a phase transition of many-body system, which is characterized by discontinuous change of system entropy or its derivatives in parameter space. Instead the dynamical behavior of a PT or CPT system takes gradual transitions across the exceptional points, and the nature of such transitions is the passage of the supermodes between their imaginary and real eigenvalues.  

Like beam-splitter arrays \cite{b1,b2,b3} as an important ingredient for linear optical quantum information processing \cite{linear}, 
our concerned CPT systems, also as a type of development from photonic molecule structures \cite{ml1,ml2}, could be suitable for the similar and other applications. Here we have demonstrated the evolution of two sets of important quantities, the cavity photon numbers and the correlation functions (average contrasts). The unnormalized correlation 
functions $\langle\hat{c}_i^\dagger\hat{c}_j\rangle$ from the latter constitute the correlation matrix or covariance matrix for a CPT system. Together with the photon numbers and average cavity modes $\langle\hat{c}_i\rangle$, these quantities fully depict the Gaussian states of such linear quantum systems \cite{gaussian, gaussian2}. 

In the above discussion we have assumed a constant amplification rate to neglect the gain saturation for not too strong cavity fields. When the cavity field intensities are sufficiently large, the amplification rate will be reduced with the growing cavity field intensities $I_i$ proportional to $\langle \hat{a}_i^\dagger\hat{a}_i\rangle$, i.e. it becomes $\kappa/(1+I_i/I_s)$ where $I_s$ is the saturation intensity. This will explicitly break the CPT symmetry of the concerned systems due to the loss of the balanced amplification and dissipation for the neighboring units. The proper operation of the systems should be under the condition 
$I_i/I_s\ll 1$. This condition of weak cavity fields compared with the saturation value necessitates the consideration of quantum noise effect discussed below. 

\section{influence of quantum noises on light transportation}
Due to CPT symmetry, the dynamical evolution matrix $e^{-i[M]t}$ in Eq. (\ref{linear}) is a symmetric one. 
If the systems are only described in terms of the non-Hermitian model in Eq. (\ref{nh}), one consequence of the symmetry is reciprocal light transportation. The light input by an external driving field into one amplification microcavity and out of another dissipation one would have the same intensity as in the reverse process (into the latter and out of the former). The violation of this property, known as optical isolation, from the saturation nonlinearity of the gain process in a PT system has been experimentally demonstrated recently \cite{ex5}. Here we will illustrate a similar phenomenon of different nature. 

The amplification and dissipation of the cavity modes are inevitably accompanied by the associated quantum noises. Generally these quantum noises are neglected in all studies of the similar systems except in \cite{arga, bhe}. The quantum noises in the CPT systems behave as the external random driving fields in Eq. (\ref{dynamics}), in addition to the coherent one to implement the light transportation. In a PT system the noise accompanying the gain process makes extra contribution to the photon numbers of the coupling modes \cite{arga}. For the CPT structure generalized from PT symmetry, there are more cavities under gain noise, and these noises act independently but affect the photon numbers in all coupled cavities. In an actual light transportation, therefore, the amplification quantum noises as the external random drives should also be considered in a strict sense. Moreover, the noise driving field intensities are independent of the coherent one, and are simply determined by the fixed amplification/dissipation rate $\kappa$. Though two coherent driving fields with the same intensity transmit symmetrically along the mutually reverse paths between an amplification and a dissipation cavity, the quantum noises of amplification add asymmetric contributions to this situation, leading to the violation of reciprocal light transportation from the non-Hermitian model.   
  
We still work with the system of $N=6$ for illustration. The light transportation can be between the first amplification cavity and the third dissipation cavity as an example. The non-reciprocal light transportation under the quantum noises can be understood from the response of the system to the external driving fields. In the non-Hermitian model, the reciprocal transportation from a coherent driving field is due to the symmetric matrix elements $[e^{-i[M](t-\tau)}]_{16}= [e^{-i[M](t-\tau)}]_{61}$ in Eq. (\ref{linear}). The amplification noises effect on the first gain cavity through the matrix elements $[e^{-i[M](t-\tau)}]_{11}$, $[e^{-i[M](t-\tau)}]_{13}$ and $[e^{-i[M](t-\tau)}]_{15}$, while drive the third loss cavity through the elements $[e^{-i[M](t-\tau)}]_{61}$, $[e^{-i[M](t-\tau)}]_{63}$ and $[e^{-i[M](t-\tau)}]_{65}$. These two sets of matrix elements have no symmetric relation. 

In Fig. \ref{fig: 6} we illustrate the difference in the mutually reverse light transportations for the three regimes across the exceptional points. These differences in the cavity photon numbers are proportional to the difference of the measured light powers. 
Here we also compare the results with those obtained with the non-Hermitian model for a transmitted light of moderate intensity,  showing that, in some of the regimes, the corrections from the quantum noises can be even more significant than the transmitted light itself. Similarly the quantum noises also affect the correlation functions we have discussed before.
The quantum noise effects will become more prominent in the CPT systems of higher cavity number $N$. Since the background noises act as the random driving fields with the intensity proportional to $\sqrt{\kappa}$ [see Eq. (\ref{h})], their effects will be overshadowed by a coherent one of sufficiently high intensity $E\gg \kappa$ (this value should go up with the cavity number $N$). However, the cavity field intensities should be limited to avoid the significant gain saturation that breaks the balanced gain and loss between the neighboring units. The amplification noises could contribute to considerable amount of the transported photons in addition to those from a not too strong coherent driving field, and should be taken into account in the general performance of the systems. 

The non-reciprocal light transportation is an appropriate phenomenon to show the non-equivalence between the non-Hermitian model in 
Eq. (\ref{nh}) and the more realistic description of the similar systems with Eq. (\ref{h}). Considering the quantum noise effects, the concerned systems do not exactly possess PT or generalized CPT symmetry. However, one of the the most important features for the systems is the dynamical transitions across the exceptional points. The existence of the exceptional points can be regarded as the consequence of a ``partial symmetry'', i.e. the PT or CPT symmetry of the dynamical matrix $[M]$ in equations of motion.

\section{conclusion}
We have studied a generalized structure from the widely explored PT symmetric optical couplers. Due to the CPT symmetry of the dynamical matrix of such systems, they possess multiple thresholds (exceptional points) for the transitions of system dynamics. 
The dynamics of an example system in the regimes separated by its exceptional points is illustrated with the evolution of cavity photon numbers and correlation functions. We find that the exceptional points of the CPT systems differ from the transitional points of many-body systems by nature; no clear-cut boundary exists for the transitions of physical quantities across them. On the other hand, we demonstrate the importance of the quantum noises with their effect on light transportation. The quantum noise effects should be considered when the external coherent driving fields may not be very strong. The understanding of the dynamics of such systems is meaningful to their potential applications.

\section*{Acknowledgement}

M.X. acknowledges funding support in part from NBRPC (Grant No. 2012CB921804) and NSFC (Nos. 61435007 \& 11321063).

\end{document}